\documentclass[tradiabstract]{aa} 

\usepackage{xcolor}

\def\bibfiles{grana_mag}
\def\aareferences{\bibliographystyle{aabib}
                 \bibliography{aajour,\bibfiles}}

\usepackage{natbib}
\usepackage{graphicx}
\usepackage{color}

\begin{document}

\title{Simulated interaction of MHD shock waves with a complex network-like region}

\titlerunning{Simulations of MHD shocks}

\author{I.C. Santamaria\inst{1,2}, E. Khomenko\inst{1,3,4}, M. Collados\inst{1,3}, A. de Vicente\inst{1,3}}
\authorrunning{I.C. Santamaria et al.}

\institute{Instituto de Astrof\'{\i}sica de Canarias, 38205 La Laguna, Tenerife, Spain
\and Centre for mathematical Plasma Astrophysics, Department of Mathematics, KU Leuven, Celestijnenlaan 200B, B-3001 Leuven, Belgium
\and Departamento de Astrof\'{\i}sica, Universidad de La Laguna, 38205, La Laguna, Tenerife, Spain 
\and Main Astronomical Observatory, NAS, 03680, Kyiv, Ukraine}

\date{Received; Accepted }

\abstract {We provide estimates of the wave energy reaching the solar chromosphere and corona in a network-like magnetic field topology, including a coronal null point. The waves are excited by an instantaneous strong subphotospheric source and propagate through the subphotosphere, photosphere, chromosphere, transition region, and corona with the plasma beta and other atmospheric parameters varying by several orders of magnitude. We compare two regimes of the wave propagation: a linear  and nonlinear regime. While the amount of energy reaching the corona is similar in both regimes, this energy is transmitted at different frequencies. In both cases the dominant periods of waves at each height strongly depend on the local magnetic field topology, but this distribution is only in accordance with observations in the nonlinear case.}

\keywords{Sun: magnetic fields; Sun: oscillations; Sun: photosphere; Sun: chromosphere; Sun: MHD waves}

\maketitle

\section{Introduction}

Magnetohydrodynamic (MHD) waves are stochastically excited below the photosphere as a result of plasma motions in the convection zone \citep{Goldreich+Keeley1977, Balmforth1992, Nordlund+Stein2001}. The magnetic field concentrations embedded in the intergranular lanes suffer footpoint motions, which also generate waves in these structures that propagate through the solar atmosphere \citep{Hasan+etal2000, Kato+etal2011} carrying with them magnetic and acoustic energy. This energy can be dissipated into heat in the upper chromosphere and corona via nonlinear events and associated microphysics. As a consequence, MHD waves are believed to be important contributors to chromospheric and coronal heating \citep[see][for a recent review]{Arregui2015}.

Of particularly interest are the questions of what amount of wave energy excited in subphotospheric layers reaches the chromosphere and corona and how the wave modes reaching there can be dissipated to convert their energy into heat, i.e., how the solar atmospheric layers are magnetically and energetically connected by means of waves. Only a few numerical studies address this question by simulating waves in the domain covering from the photosphere to the corona \citep{Fedun+etal2011b, Santamaria+etal2015}. In their 2D simulations of high-frequency wave propagation, \citet{Fedun+etal2011b} find that magneto-acoustic waves in a wide range of high frequencies can leak energy effectively into the corona. These authors considered an isolated magnetic flux tube that becomes vertical in the chromosphere. \citet{Santamaria+etal2015} show the energy propagation in the linear regime by waves excited below the photosphere from different driving mechanisms that propagate in a complex magnetic field configuration. They use drivers with more realistic periods of 3-5 minutes and conclude that the waves get into the corona more preferably inside the vertical flux tubes, being the energy is of acoustic nature.

Here we extend the latter work with the aim of studying the influence of nonlinearities in the wave behavior and energy transport to the chromosphere and corona. To that purpose, we perform simulations of waves driven by an instantaneous pressure pulse located below the photosphere with an amplitude that is sufficiently large to develop nonlinearities. This model is compared with an identical simulation studied previously in \citet{Santamaria+etal2015}, except for the strength of the driving pulse causing a linear wave propagation.

\begin{figure*} 
\centering
\includegraphics[width=5cm]{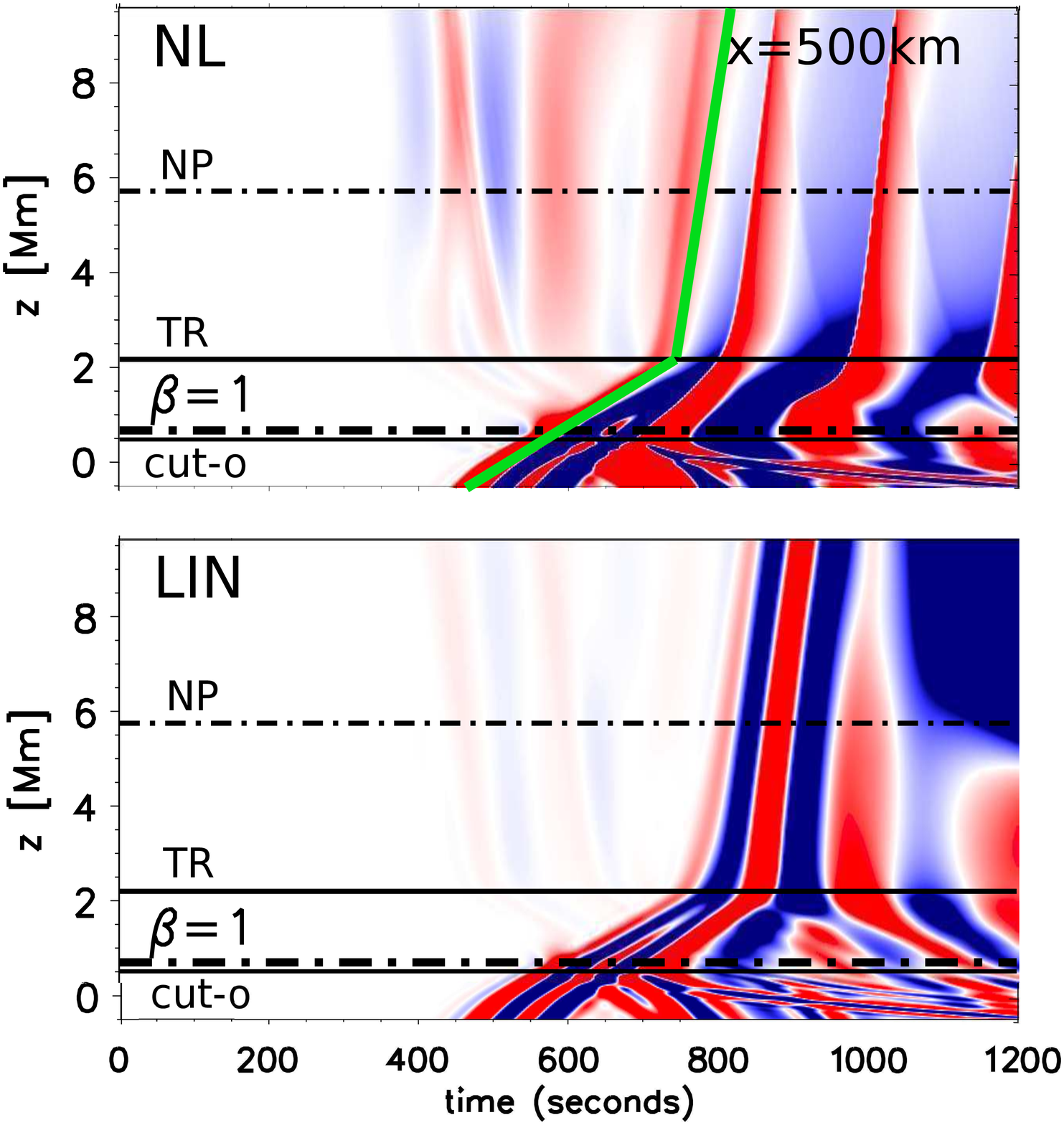}
\includegraphics[width=5cm]{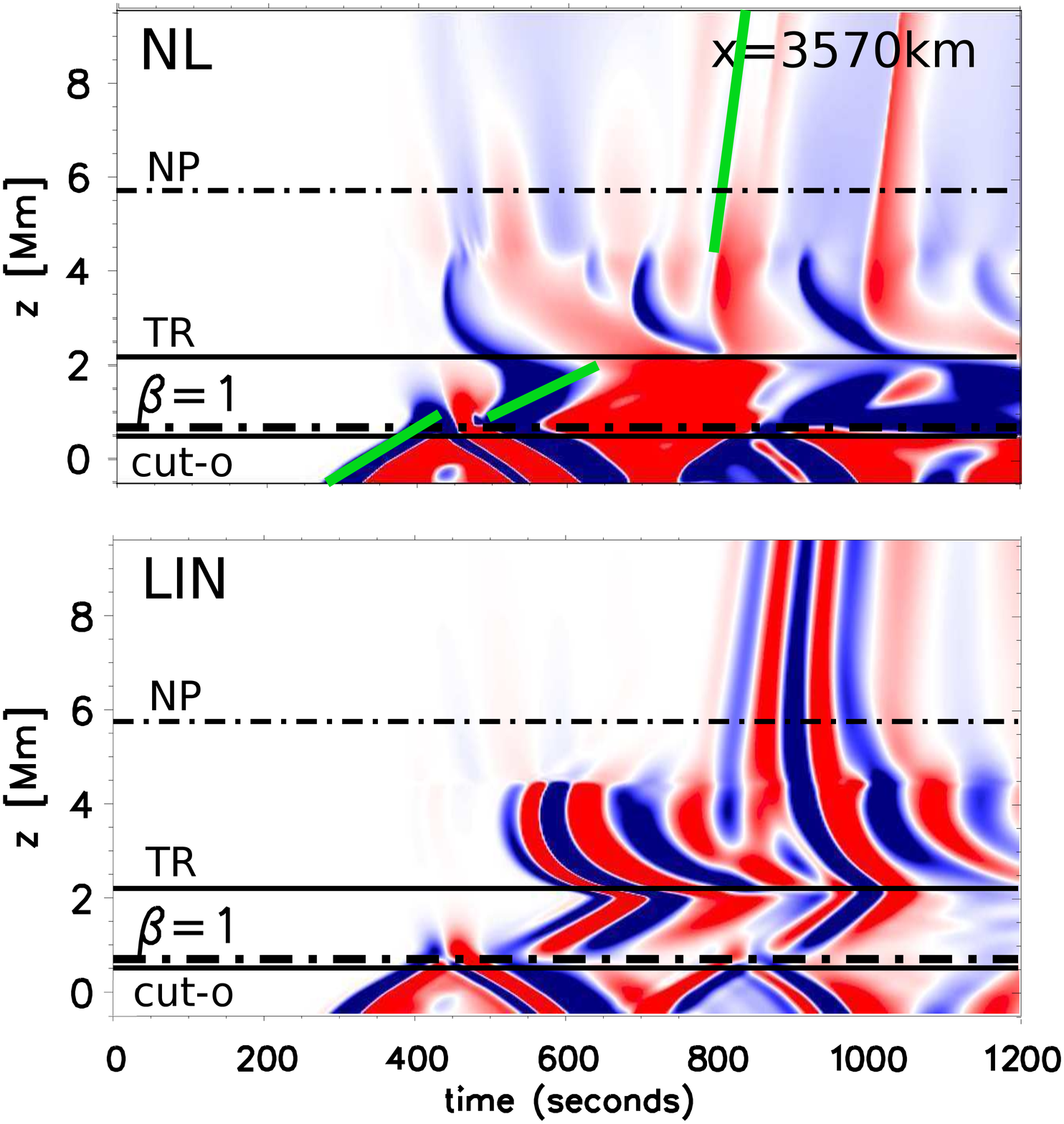} 
\includegraphics[width=5cm]{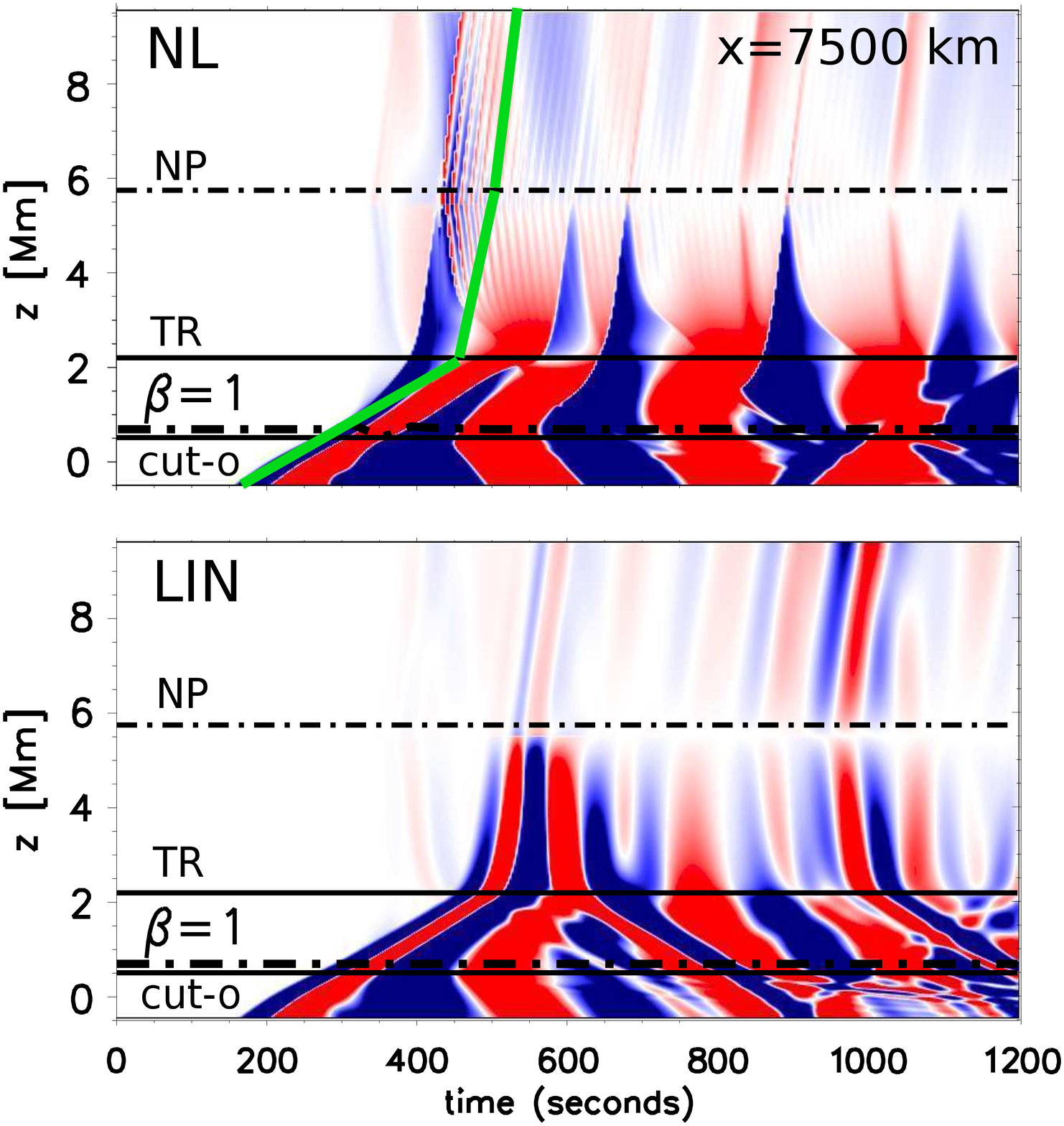}  
\includegraphics[width=1.6cm]{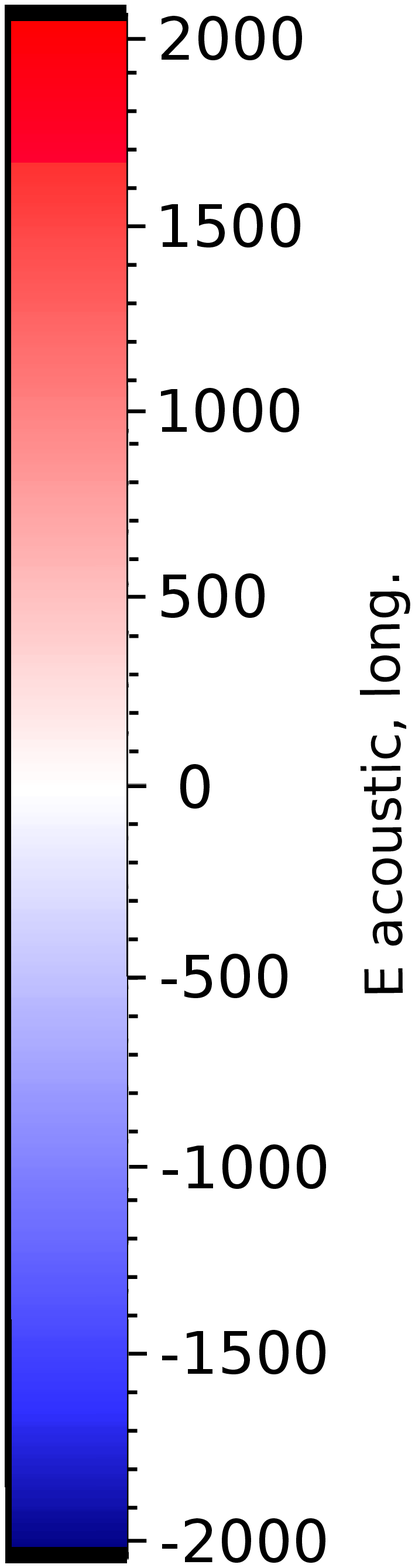}  

\caption{Perturbation of the acoustic energy flux proxy in the direction parallel to the magnetic field for three different horizontal positions: (\textit{left}) $x=500$ km, inside the vertical flux tubes; (\textit{middle}) $x=3750$ km, in the middle of the arcades where the magnetic field is mostly horizontal; and (\textit{right}) $x=7500$ km, in the middle of the domain where the null point is located. Upper row: nonlinear case, bottom row: linear case. The energy proxy is given in $\sqrt{erg/cm^{2}s}$. The plots were scaled to the corresponding amplitude of the pressure driver so that the colors in each figure can be directly compared. Overplotted green lines in the upper panels show the slopes on the linear fronts to compare both regimes more directly. }\label{fig:fl}
\end{figure*}
\begin{figure*} 
\centering
\includegraphics[width=5cm]{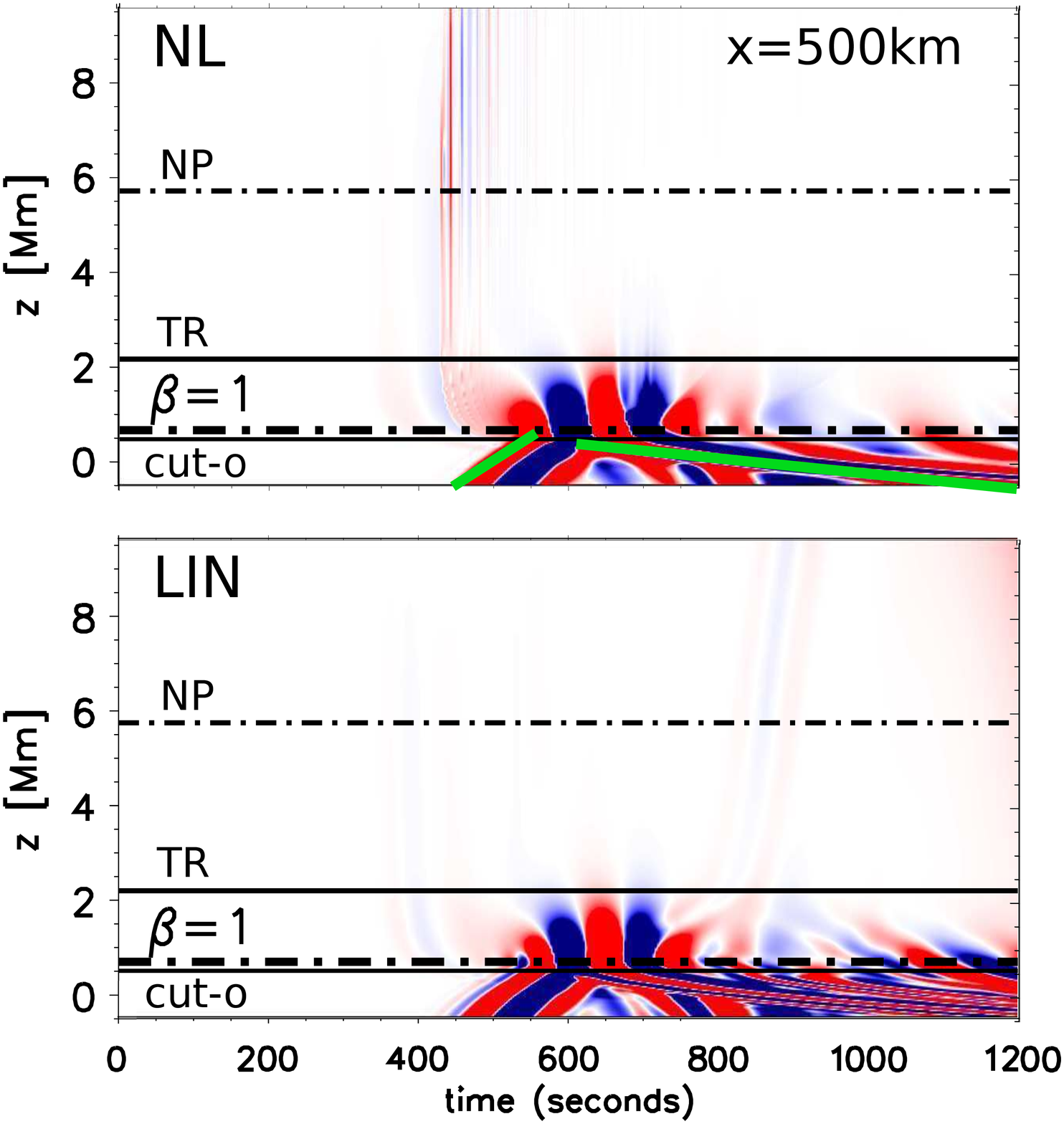} 
\includegraphics[width=5cm]{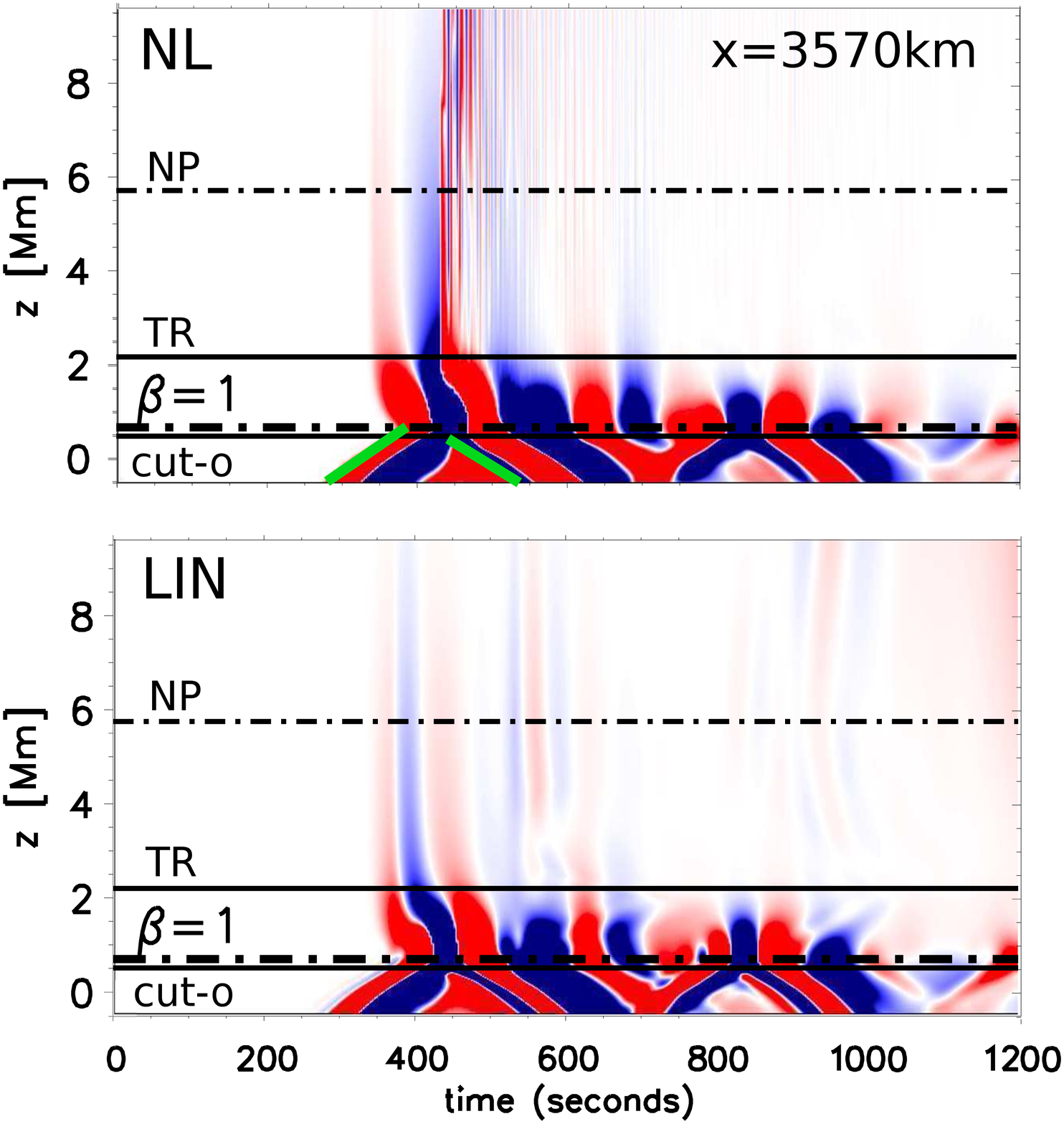}
\includegraphics[width=5cm]{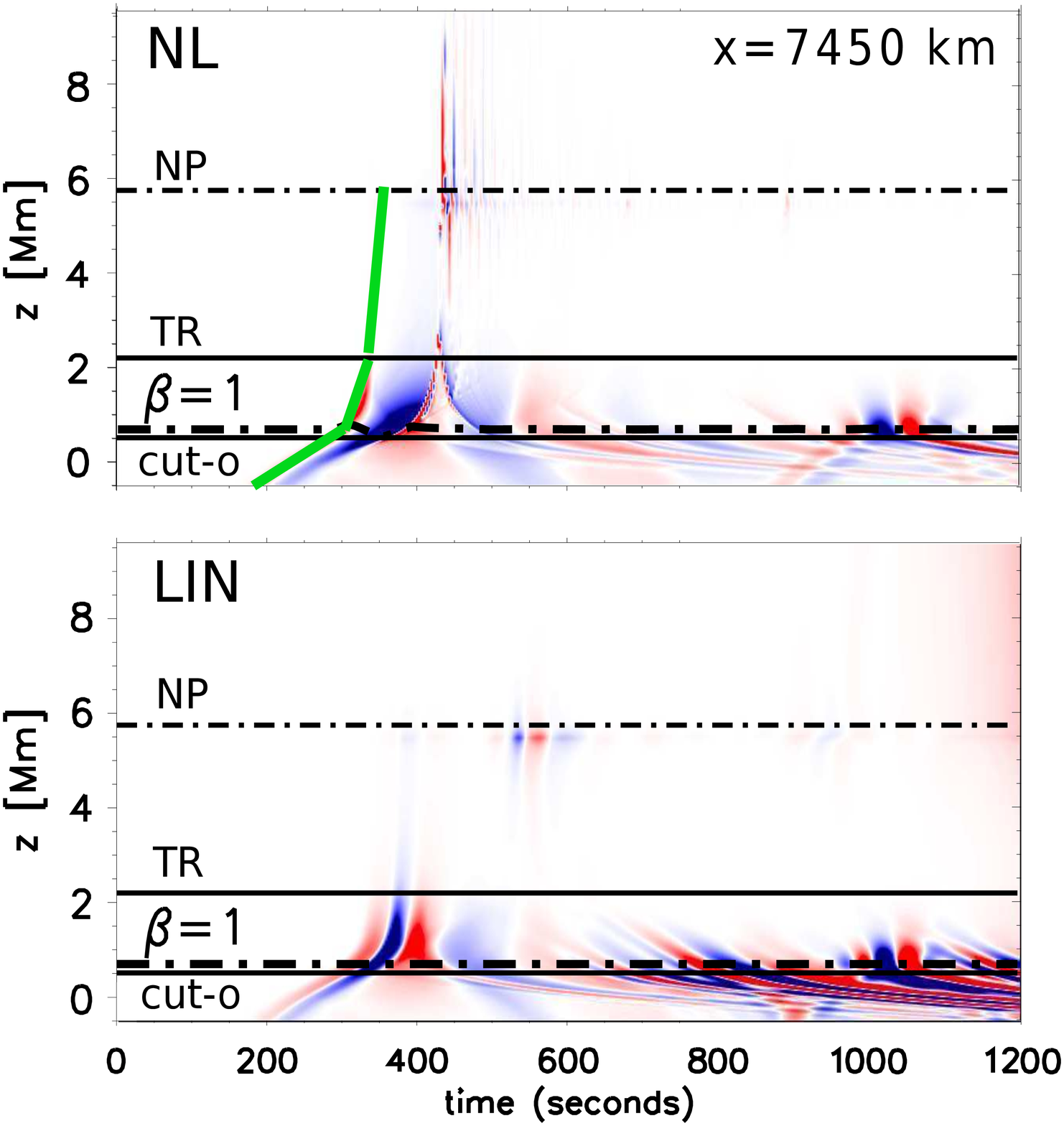} 
\includegraphics[width=1.6cm]{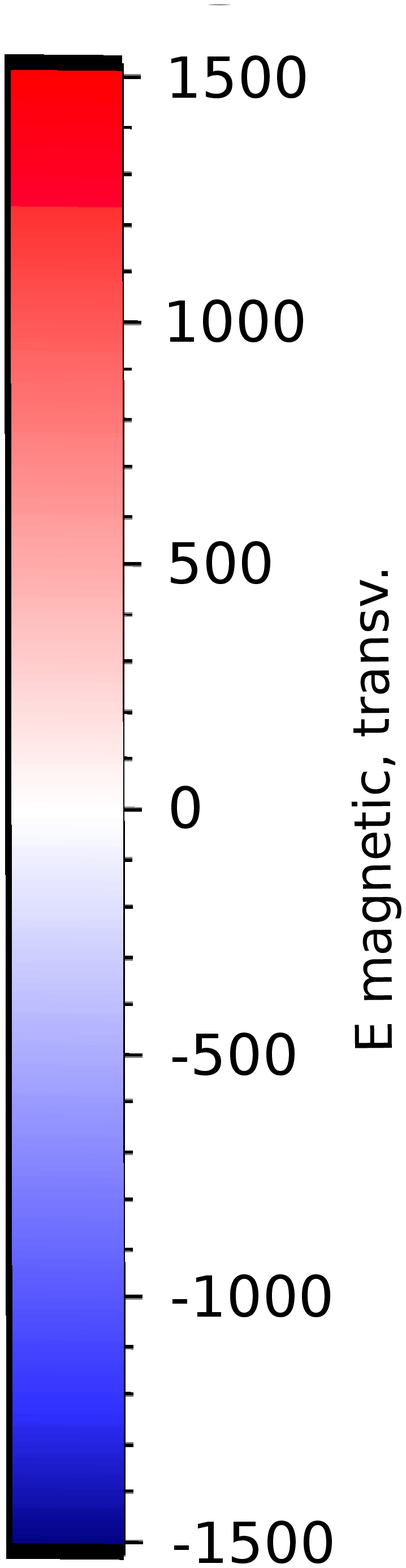}  
 
\caption{Perturbation of the magnetic energy flux proxy in the direction transverse to the magnetic field. The format and units of the figure is same as Fig. \ref{fig:fl}.}\label{fig:ft}
\end{figure*}

\section{Numerical method}
We solve in 2D the ideal magnetohydrodynamic equations of conservation of mass, momentum, energy, and the induction equation for the magnetic field,
\begin{equation} \label{eq:den}
\frac{\partial\rho}{\partial t}+\nabla \cdot(\rho{\bf v})= \Big [ \frac{\partial\rho}{\partial t} \Big ]_{\rm diff} \,, 
\end{equation}
\begin{equation} \label{eq:mom}
\frac{\partial (\rho{\bf v})}{\partial t}+\nabla\Big [\rho{\bf vv}+\Big (p+\frac{{\bf B}^2}{2 \mu_0}\Big ){\bf I}-\frac{{\bf B}{\bf B}}{\mu_0}\Big ]=\rho{\bf g}  + \Big [ \frac{\partial (\rho{\bf v})}{\partial t} \Big ]_{\rm diff}    \,,
\end{equation}
\begin{equation} \label{eq:ei}
\frac{\partial p}{\partial t}+ (\textbf{v} \cdot \nabla) p + \gamma p (\nabla \cdot \textbf{v}) = \Big [\frac{\partial p}{\partial t} \Big]_{\rm diff} \,,
\end{equation}
\begin{equation} \label{eq:ind}
\frac{\partial {\bf B}}{\partial t}=\nabla\times ({\bf v} \times {\bf B}) + \Big [ \frac{\partial {\bf B}}{\partial t}\Big ]_{\rm diff} \,,
\end{equation}
\noindent where all the notations are standard. We use an ideal equation of state and $\gamma = 5/3$. The terms with subscript ``diff'' are the artificial hyper-diffusive terms required for the numerical stability of the code \citep[see][]{Felipe+etal2010}. After removing the equilibrium condition, we solve these nonlinear equations by means of the code {\sc mancha,} which is described in detail in \citet{Khomenko+Collados2006} and \citet{Felipe+etal2010}. The numerical code solves the nonlinear equations for perturbations. Nevertheless, in the simulations where the linear regime is desired, the amplitude of the perturbations can be kept small enough to avoid nonlinearities. 

\subsection{Simulation setup}
\begin{figure*}
\centering
\includegraphics[width=14cm]{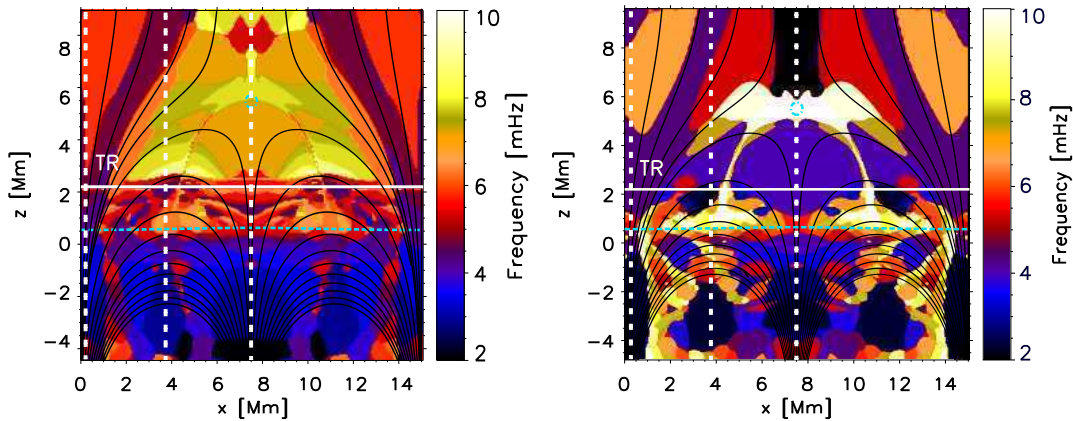}
\caption{Spatial distribution of the dominant frequencies of the vertically propagating waves for a maximum frequency of 10 mHz for \textit{(left)} linear and \textit{(right)} nonlinear regimes. Dashed blue lines denote $\beta=1$ contours, white solid horizontal lines indicate the transition region and black solid lines indicate magnetic field lines. The vertical white dashed lines indicate the vertical slices for which the time-height diagrams are shown in Figs. \ref{fig:fl} and \ref{fig:ft}.} 
 \label{fig:periods_small}
\end{figure*}

We make use of the atmospheric models described in \citet{Santamaria+etal2015} to perform these simulations. The magnetic field is potential, which is composed of two vertical flux tubes of the same polarity and separated by an arcade-shaped field (see Fig. \ref{fig:periods_small}). This magnetic field configuration is superimposed on a plane-parallel atmosphere with vertical stratifications in pressure, density, and temperature. The two simulations analyzed here are identical in all parameters except for the strength of the driving pressure pulse. This pulse is instantaneous and is located at the base of the arcades in the middle of the domain
\begin{eqnarray} \label{eq:pulse_p}
\frac{\delta p_{1}}{p_0} &=& A \gamma \exp \Big [-\Big( \frac{(x-x_{0})^{2}}{2\sigma_{x}^{2}} + \frac{(z-z_{0})^{2}}{2\sigma_{z}^{2}} \Big) \Big] \\
\frac{\delta \rho_{1}}{\rho_0} &=& A \exp \Big [-\Big( \frac{(x-x_{0})^{2}}{2\sigma_{x}^{2}} + \frac{(z-z_{0})^{2}}{2\sigma_{z}^{2}} \Big) \Big]
,\end{eqnarray}
\noindent where $A$ is the relative amplitude of the pulse, which is $A=0.2$ in the nonlinear and $A=10^{-5}$ in the linear case, $[x_0,z_0]=[7500,-3500 ]$ km are the coordinates at which the pulse is located, and $\sigma_{x}$=$\sigma_{z}$=700 km are the widths of the two-dimensional Gaussian profile. Such a pulse generates a superposition of waves in a broad range of frequencies.

To avoid spurious reflections of waves at the vertical boundaries of the domain, we add two perfectly matched layers \citep[PML;][]{Berenger1994} of 20 (bottom) and 40 (top) grid points; see \citet{Felipe+etal2010} for details. 
Periodic boundary conditions are used in the horizontal direction. The duration of the simulations is 3000 sec.

\section{Wave behavior}

We proceed by calculating the acoustic and magnetic wave energy flux proxies, given by the longitudinal and transverse velocities multiplied by $\sqrt{\rho_0 c_{s}}$ and $\sqrt{\rho_0 v_{a}}$, respectively, where $c_{s}$ is the sound speed and $v_{a}$ refers to the Alfv\'{e}n speed. These quantities give us an approximation of acoustic and magnetic energy fluxes propagating in the atmosphere, along and transverse to the magnetic field lines, respectively. 

Figs. \ref{fig:fl} and \ref{fig:ft} show time-height diagrams for acoustic and magnetic energy flux proxies, respectively, at three representative horizontal locations of the domain, which are indicated in the figure for the nonlinear (top) and linear (bottom) cases. The energy flux proxies were normalized to the corresponding amplitude of the pulse to facilitate a direct comparison of the figures. Since the nonlinear phenomena are more pronounced in the upper atmosphere, we first focus on the region above 0 km. 
We only show half of the simulation time because the energy proxies start to decrease considerably after $t=1500$ s as a result of the energy escape from the domain, and no additional information is obtained from subsequent time steps. 

In Fig. \ref{fig:fl}, which shows the acoustic energy flux proxy, it can be appreciated that the linear and nonlinear wave fronts differ significantly. Inside the vertical flux tubes (left panels), the acoustic energy reaches the corona faster in the nonlinear regime. The larger propagation velocity is a direct consequence of the nonlinear wave behavior produced by the larger velocities generated by the stronger driver. This effect can be identified by the more vertical orientation of the wave fronts in the time-height diagrams corresponding to the nonlinear regime (clearly noticeable in the chromosphere and above). The acoustic energy reaching the corona at the time around $t\sim 400$ s is clearly larger in the nonlinear case. After 700 seconds, the acoustic energy flux fronts reaching the corona from the lower layers of the flux tube are steeper in the nonlinear regime.\

At the middle of the arcades (central panels of Fig. \ref{fig:fl}), where the magnetic field is almost horizontal, the energy flux also reaches the corona faster in the nonlinear case. This is because the selected location ($x=3750$ km) for the time-height diagram crosses the expanding flux tube above 4 Mm, where where propagating acoustic waves are aligned in the field (see dashed vertical lines in Fig. \ref{fig:periods_small}). As a consequence, the acoustic energy flux appears in the corona earlier in the nonlinear case, due to the larger wave propagating speeds. At the same time, the region at the middle of the arcades is strongly influenced by the null point. The movie of the simulations (see Appendix \ref{app}) shows how waves reaching the null point are resent outward again in all directions. The waves coming from the null point together with those coming from the flux tubes and lower layers, lead to a superposition of waves, making the wave fronts cancel each other out. This can be appreciated more clearly in the linear regime, where there is a lack of acoustic energy above 4 Mm.

The right panels of Fig. \ref{fig:fl} show the acoustic energy flux proxy at the middle of the domain where the null point is located. The acoustic energy flux propagation again differs between the linear and nonlinear regimes. The wave fronts in the nonlinear case are sharper as waves steepen to shocks. At the time interval between 400 and 500 s, we observe a curious phenomenon of a high-frequency wave propagation just below and above the null point. This phenomenon is only present in the nonlinear regime. The movie of the simulations (see Figure \ref{movie:nonlin} in Appendix \ref{app}) shows that these waves are triggered strongly outward from the null point. They propagate until the end of the simulation time, but their amplitude decreases as they escape from the domain through the upper boundary. Below the null point we see sharp fronts in the nonlinear case. These are, again, hydrodynamic shock waves propagating from the chromosphere to the corona in a rather vertical field. 

Fig. \ref{fig:ft} shows the magnetic energy flux proxies for both linear and nonlinear cases. The time-height diagrams at the same three different vertical slices of the domain show that the magnetic energy flux behavior is very similar in both cases, except for the first strong front reaching the upper corona in the nonlinear case. The similar behavior of waves in the two different regimes suggests that the magnetic energy is less influenced by nonlinearities, which is expected because the wave amplitudes are sub-Alfv\'enic. At the middle of the arcades (central panels) the wave front reaching the corona is composed of high-frequency oscillations, as mentioned above. We defer the discussion of these oscillations to a separate paper.
To analyze the magnetic energy flux proxy close to the null point, we moved one grid point to the left (x=7450 km instead of 7500 km),= because in the middle of the domain the energy flux is almost zero (right panels of Fig. \ref{fig:ft}). Most of the energy flux stays around the equipartition layer owing to the refraction of fast-magnetic waves, as seen for the other locations in the domain. 

As a result, apart from the (expected) larger velocities and shock formation in the nonlinear regime, a most important result is that the interaction with the null point is different between the two regimes. While the nonlinear simulation creates high-frequency waves around the null point, the linear case results in a different interference pattern below the transition region. The fraction of energy reaching the chromosphere and corona is, however, similar for both regimes.

\section{Frequency distribution}

The pressure pulse produces waves in a broad range of frequencies. The dependence between the dominant wave frequency and the magnetic topology is shown in Fig. \ref{fig:periods_small}. This figure compares the linear case left panel with the nonlinear case right panel. In the nonlinear case the waves around the null point have a frequency of 80 mHz, while the frequency does not exceed 20 mHz in the rest of the domain. For purposes of better visualization and comparison between linear and nonlinear propagation regimes, we limited the maximum frequency to 10 mHz in Fig. \ref{fig:periods_small}. The high-frequency region appearing owing to nonlinear effects around the null point will be analyzed in detail in forthcoming works.

According to the left panel of Fig. \ref{fig:periods_small}, in the linear regime we observe 3-4 mHz oscillations in the photosphere, which is below equipartition layer at all locations except for some regions at the interface between the flux tubes and the arcades. In the chromosphere we observe 5-6 mHz oscillations in vertical (or close to vertical) magnetic fields, while higher frequencies are found in more inclined and horizontal magnetic field regions. 

In the nonlinear regime (right panel of Fig. \ref{fig:periods_small},) this picture changes. We observe 4 mHz oscillations reaching the corona along the external part of the flux tubes, where the magnetic field is more inclined, while at the internal part of the flux tubes, oscillations of 6 mHz dominate. This distribution resembles that observed in a solar plage or in the network \citep{Deubner+Fleck1990, Lites+etal1993, Vecchio+etal2007, deWijn+etal2009, Khomenko+CalvoSantamaria2013}, unlike in the linear case where only oscillations with frequencies around 5 mHz are transmitted through the transition region to the corona at the locations around the flux tubes.

Outside the vertical flux tubes, in the corona, frequencies in the range 7--10 mHz are found in the linear case (yellow regions above 2 Mm at the left panel). In the nonlinear regime, waves with these high frequencies are refracted at the transition region propagating downward along the magnetic canopy (yellow regions around 1 Mm height at the right panel). The remaining oscillations above the arcades and below the null point have frequencies of 4 mHz.
Therefore, we conclude that only the nonlinear case adequately describes the observed frequency distribution.

\section{Discussion and conclusions}

In this work, we have analyzed the nonlinear behavior of MHD waves by comparing two simulations that only differ by the strength of the driving pulse. We find that the time-height diagrams for the acoustic energy differ significantly, while this is not the case for the magnetic energy because the amplitudes of the waves are sub-Alfv\'enic, but not sub-sonic. Despite the larger propagation speed and the presence of shock waves in the nonlinear simulation that change the appearance of the time-height diagrams, we find that the general picture of the mode propagation and fast wave refraction is similar in both regimes. However, the difference in the wave propagation speeds results in a different interference pattern of waves at heights between the equipartition region (around 1 Mm) and the transition region (above 2 Mm). The interaction of the waves with the null point is also different, producing an apparent high-frequency pattern in the nonlinear regime. 

While the fraction of energy that reaches the corona is similar in both regimes, we find that it is transported at different frequencies. The analysis of the relation between the topology of the magnetic field and the dominant frequency of waves in both linear and nonlinear regimes shows that the nonlinear wave propagation is more realistic than the linear wave propagation. In the nonlinear case we obtain three minute oscillations at locations with more vertical field and four to five minute oscillations at locations with a more inclined field around flux tubes. Such frequency distribution, also seen in simulations by \citet{Heggland+etal2011}, resembles plage and network areas.
In the corona above the arcades with a horizontal magnetic field, we find high-frequency oscillations in the linear regime, but 5-min oscillations in the nonlinear regime, which is again more realistic.

\begin{acknowledgements}
This work is partially supported by the Spanish Ministry of Science through projects AYA2010-18029 and AYA2011-24808.
This work contributes to the deliverables identified in FP7 European Research Council grant agreement 277829, ``Magnetic Connectivity through the Solar Partially Ionized Atmosphere''.
\end{acknowledgements}

\aareferences
\newpage
\begin{appendix} 
\section{Available movies}
\label{app}
This Appendix aims to briefly explain what is shown in the available online movies.
\begin{figure}[!h]
\includegraphics[width=10cm]{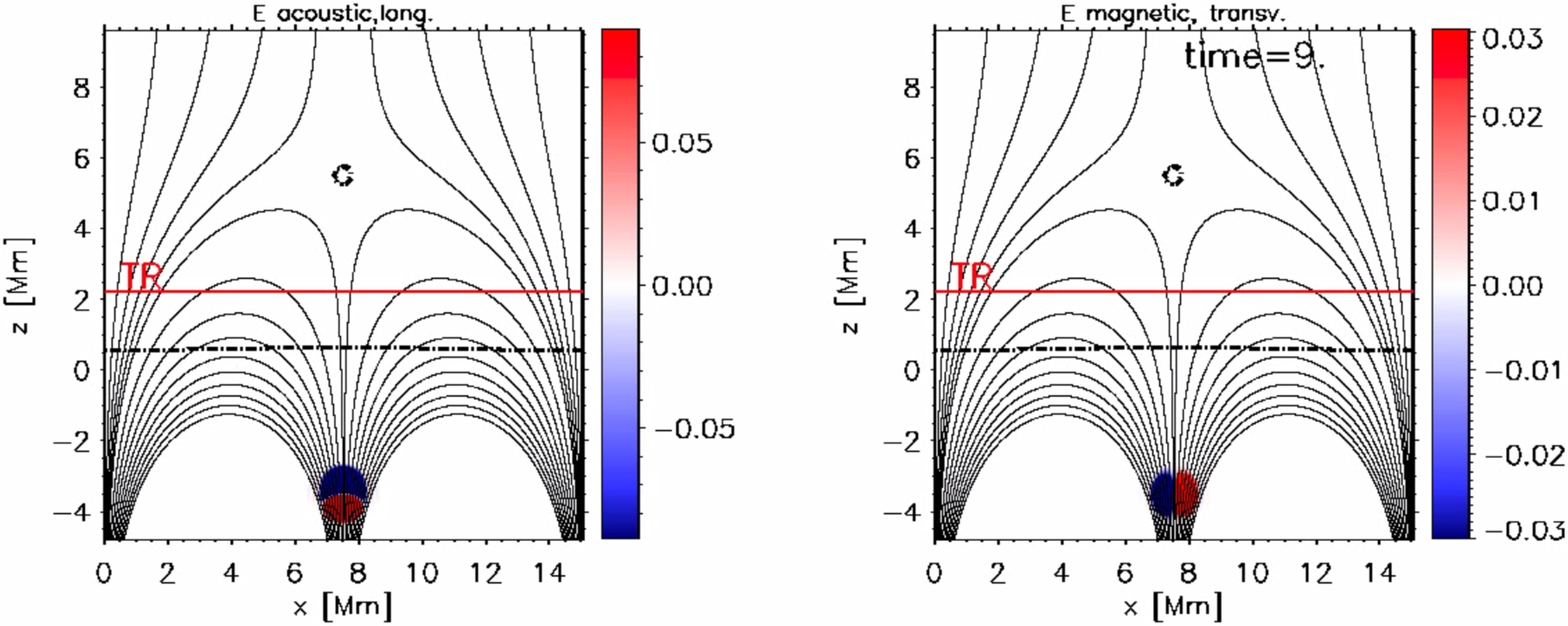}
\caption{Attached movie shows the linear wave propagation by means of acoustic (left panel) and magnetic (right panel) energy flux proxies. These quantities are the same as shown in Figures \ref{fig:fl} and \ref{fig:ft} and the units are given in $\sqrt{erg/cm^{2}s}$ . The equipartition layers ($\beta$ = 1 layer) are indicated with black dashed lines, while the transition region is shown with a red solid line. }
\label{movie:lin}
\end{figure}
\begin{figure}[!h] 
\includegraphics[width=10cm]{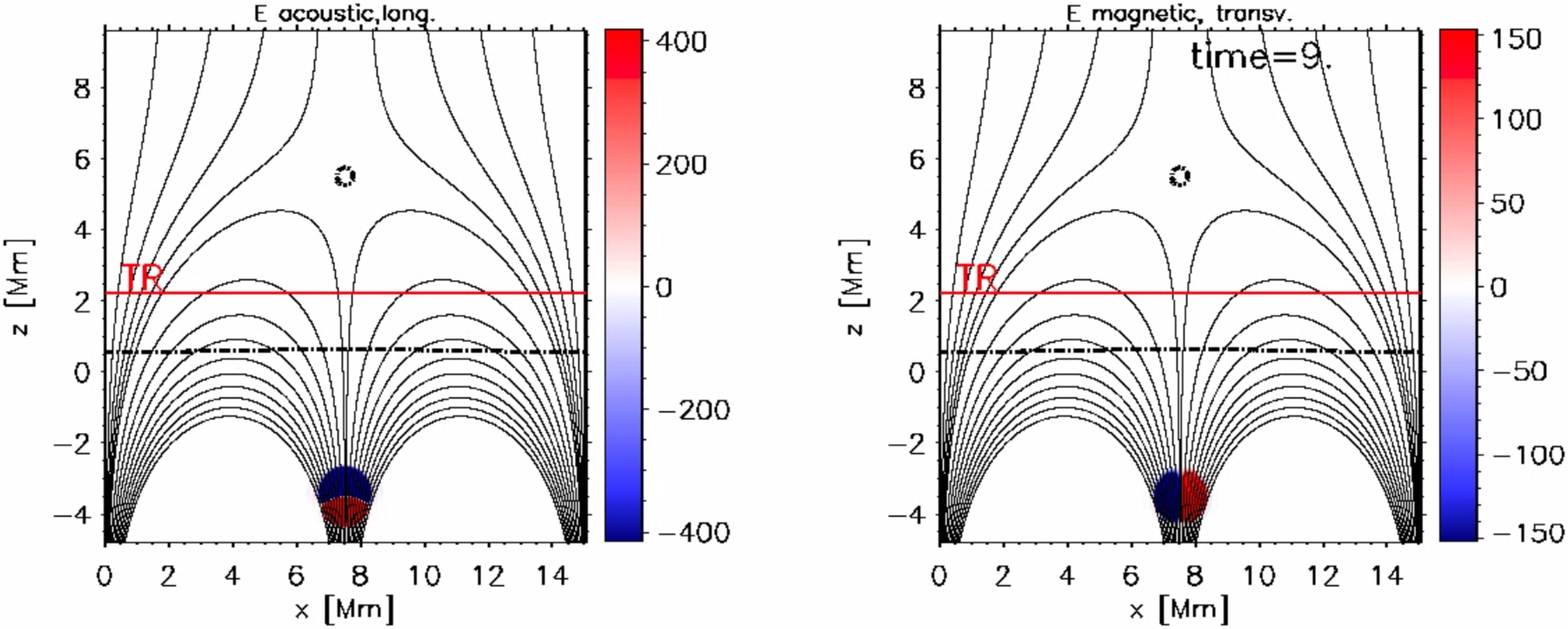}
\caption{Attached movie shows the nonlinear wave propagation by means of acoustic (left panel) and magnetic (right panel) energy flux proxies. These quantities are the same as shown in Figures \ref{fig:fl} and \ref{fig:ft} and the units are given in $\sqrt{erg/cm^{2}s}$ . The equipartition layers ($\beta$ = 1 layer) are indicated with black dashed lines, while the transition region is shown with a red solid line.}
\label{movie:nonlin}
\end{figure}

\end{appendix}

\begin{thebibliography}{}

\bibitem[\protect\astroncite{{Arregui}}{2015}]{Arregui2015}
{Arregui}, I. 2015, ArXiv e-prints

\bibitem[\protect\astroncite{Balmforth}{1992}]{Balmforth1992}
Balmforth, N.~J. 1992, MNRAS, 255, 639

\bibitem[\protect\astroncite{Berenger}{1994}]{Berenger1994}
Berenger, J.~P. 1994, J.\ Comp.\ Phys., 114, 185

\bibitem[\protect\astroncite{{de Wijn} et~al.}{2009}]{deWijn+etal2009}
{de Wijn}, A.~G., {McIntosh}, S.~W., {De Pontieu}, B. 2009, ApJ, 702, L168

\bibitem[\protect\astroncite{{Deubner} \& {Fleck}}{1990}]{Deubner+Fleck1990}
{Deubner}, F.-L., {Fleck}, B. 1990, A\&A, 228, 506

\bibitem[\protect\astroncite{{Fedun} et~al.}{2011}]{Fedun+etal2011b}
{Fedun}, V., {Verth}, G., {Jess}, D.~B., {Erd{\'e}lyi}, R. 2011, ApJ, 740, L46

\bibitem[\protect\astroncite{{Felipe} et~al.}{2010}]{Felipe+etal2010}
{Felipe}, T., {Khomenko}, E., {Collados}, M., {Beck}, C. 2010, ApJ, 722, 131

\bibitem[\protect\astroncite{Goldreich \& Keeley}{1977}]{Goldreich+Keeley1977}
Goldreich, P., Keeley, D.~A. 1977, ApJ, 211, 934

\bibitem[\protect\astroncite{{Hasan} et~al.}{2000}]{Hasan+etal2000}
{Hasan}, S.~S., {Kalkofen}, W., {van Ballegooijen}, A.~A. 2000, ApJ, 535, L67

\bibitem[\protect\astroncite{{Heggland} et~al.}{2011}]{Heggland+etal2011}
{Heggland}, L., {Hansteen}, V.~H., {De Pontieu}, B., {Carlsson}, M. 2011, ApJ,
  743, 142

\bibitem[\protect\astroncite{{Kato} et~al.}{2011}]{Kato+etal2011}
{Kato}, Y., {Steiner}, O., {Steffen}, M., {Suematsu}, Y. 2011, ApJ, 730, L24

\bibitem[\protect\astroncite{{Khomenko} \& {Calvo
  Santamaria}}{2013}]{Khomenko+CalvoSantamaria2013}
{Khomenko}, E., {Calvo Santamaria}, I. 2013, Journal of Physics Conference
  Series, 440(1), 012048

\bibitem[\protect\astroncite{Khomenko \&
  Collados}{2006}]{Khomenko+Collados2006}
Khomenko, E., Collados, M. 2006, ApJ, 653, 739

\bibitem[\protect\astroncite{{Lites} et~al.}{1993}]{Lites+etal1993}
{Lites}, B.~W., {Rutten}, R.~J., {Kalkofen}, W. 1993, ApJ, 414, 345

\bibitem[\protect\astroncite{Nordlund \& Stein}{2001}]{Nordlund+Stein2001}
Nordlund, {\AA}., Stein, R.~F. 2001, ApJ, 546, 576

\bibitem[\protect\astroncite{{Santamaria} et~al.}{2015}]{Santamaria+etal2015}
{Santamaria}, I.~C., {Khomenko}, E., {Collados}, M. 2015, \aap, 577, A70

\bibitem[\protect\astroncite{Vecchio et~al.}{2007}]{Vecchio+etal2007}
Vecchio, A., Cauzzi, G., Reardon, K.~P., Janssen, K., Rimmele, T. 2007, A\&A,
  461, L1

\end{thebibliography}
\end{document}